\documentclass{article}
\usepackage[affil-it]{authblk}
\usepackage{bm} 
\usepackage{subfigure}
\usepackage{amsthm}
\usepackage{amsmath}
\usepackage{graphicx}
\usepackage{amssymb}

\usepackage{algorithm2e}
\usepackage{natbib}
 \bibpunct[, ]{(}{)}{,}{a}{}{,}%
 \def\newblock{\ }%
\usepackage{bm}
\newtheorem{proposition} {Proposition}
\newtheorem{assumption} {Assumption}
\newtheorem{theorem} {Theorem}
\newtheorem{lemma} {Lemma}

\title{Sequential Selection for Accelerated Life Testing via Approximate Bayesian Inference}

\author[1]{Ye Chen}
\author[2]{Qiong Zhang}
\author[3]{Mingyang Li}
\author[4]{Wenjun Cai}
\affil[1]{ Department of Statistical Sciences and Operations Research, Virginia Commonwealth University}
\affil[2]{School of Mathematical and Statistical Sciences, Clemson University}
\affil[3]{Department of Industrial and Management Systems Engineering, The University of South Florida}
\affil[4]{Department of Materials Science \& Engineering, Virginia Polytechnic Institute and State University}
\date{}

\begin{document}


\maketitle

\begin{abstract}
 Accelerated life testing (ALT) is typically used to assess the reliability of material's lifetime under desired stress levels. 
Recent advances in material engineering have made a variety of material alternatives readily available. To identify the most reliable material setting with efficient experimental design, a sequential test planning strategy is preferred.  
 To guarantee a tractable statistical mechanism for information collection and update, we develop explicit model parameter update formulas via 
approximate Bayesian inference. Theories show that our explicit update formulas give consistent parameter estimates. Simulation study and a case study show that the proposed sequential selection approach can significantly improve the probability of identifying the material alternative with best reliability performance over other design approaches.
\end{abstract}


	\noindent%

\section{Introduction}

\subsection{Motivation}
Product reliability is often referred to as its ability of performing intended function under specific operating conditions. However, it might take months or years to observe a product failure under the desired operating conditions. Accelerated life test (ALT) is used to collect reliability information in a timely manner under accelerated operating conditions in the lab environment.
Then the reliability information collected can be used 
to predict the lifetime under the normal operating conditions in field environment. Typically, ALT tests $N$ experimental units due to the availability of experimental resource. The classical problem of experimental design for ALT is to allocate the  stress levels representing accelerated operating conditions
to each test unit.

Recent advances in material engineering have made a variety of material settings readily available in lab testing. 
Among those different settings, the proportions of different elements in the material and  mechanical procedures would greatly influence their reliability performance. 
Thus, the selection of material setting is often critical to the product reliability. 
In this paper, a new task for ALT is to select the material setting with the best reliability performance. To fulfill this aim, the problem of experimental design for ALT is to determine the stress levels, as well as the material setting of each test unit. As demonstrated in \cite{lee2018sequential}, sequential design is often preferable compared to one-shot designs (i.e., allocating design points for all test units at the beginning stage of the experiments) in terms of improving the efficiency of test planning. The reasons are given as follows. First,  
testing labs are typically equipped with only a limited number of testing machines (e.g., one or two). Therefore, it is 
physically impossible to conduct all $N$ experiments simultaneously. Second, efficient one-shot design relies on prior estimates of model parameter, and an accurate prior of model parameters is often difficult to obtain before conducting ALT. Particularly, this paper focuses on selecting the optimal material setting, and the advantage of the sequential test planning is to improve the efficiency in optimal decision-making.

To the best of our knowledge, there is no existing work discussing sequential design for optimal material selection under the framework of ALT. We propose a sequential selection approach to allocate experimental design settings to the test units. In each step of this sequential procedure, the experimental design setting for the new test unit is selected to maximize the expected gain on optimizing the reliability performance under a Bayesian log-normal model. 
For the computational convenience of sequential selection,
we develop explicit model parameter update formulas via 
approximate Bayesian inference. 
Theories show that our explicit update formulas give consistent parameter estimates. In the next subsection, we point out the connection of our work to literature studies.

\subsection{Related Literature}

Our paper is closely related to the literature of experimental design for ALTs, as well as the literature on sequential experimental design and learning in simulation optimization. We review state-of-the-art approaches and recent advances from both communities and point out their connections to our paper. 

The typical problem in designs for ALT is to allocate accelerated stress levels to experimental test units. The ASTM standard \citep{standard10e739} suggests balanced and equally spaced designs for ALT. Given a lower bound and an upper bound of a stress factor, 
equally spaced design points are chosen. Each design point is applied to an equal number of experimental units. This standard design is developed to reduce the variance of parameter estimates or prediction. To achieve the optimal efficiency in parameter estimation or prediction, optimum test planning strategies have been developed under different model settings, see for examples, \cite{meeker1977asymptotically,meeker2014statistical, pan2014design, king2016planning}. Those optimal design approaches work well if the substituted parameter guesses in the model are accurate. This requirement is often impractical at the early stage of the experimentation. Recently, \cite{lee2018sequential} developed a sequential Bayesian design approach for ALT to mitigate this drawback, and improve the efficiency in test planning. 
However, as noted earlier, most of existing experimental design approaches for ALT are developed to assess the reliability performance of a given product or material. In this paper, we focus on selecting the optimal material setting with the best reliability performance. The experimental design issue for this particular problem has not been discussed in the literature to the best of our knowledge. 

Selecting the optimal design among different alternatives has been well known as the ranking and selection (R\&S) problem in the simulation community, which can date back to \cite{Be54a}. In such problems, the experiment is usually under the limit of a fixed budget (for example, time, materials), and the decision-maker wants to identify the optimal design correctly as much as possible. See \cite{HoNe09} and \cite{ChFuQuRy14} for more description. For the R\&S problem, we say ``correct selection'' occurs if the selected alternative is truly the best design after the simulation budget is exhausted. The optimal budget allocation with respect to maximizing the probability of the correct selection is studied rigorously in \cite{GlJu04}. However, this optimal budget allocation requires certain knowledge of the designs and thus can not be applied directly in practice; for more details, see the discussion in \cite{ChRy2019cei}. Therefore, modern researchers prefer to allocate their budget in a sequential manner, which is more practical and computationally tractable. In such sequential allocation algorithms, the decision-maker first spends part of the budget, observes the results, then determines how to allocate the remaining budget accordingly. There are many sequential allocation algorithms that have been proposed, including expected improvement \citep[or EI; see][]{JoScWe98}, optimal computing budget allocation \citep[or OCBA; see][]{ChLiYuCh00}, indifference-zone method \citep{KiNe01}, top-two methods \citep{Ru17}. The EI-type methods also include \cite{ChBrSc10, powell2012optimal, QiKlRu17, SaSoNeSt19}. Other approaches include the reverse-engineering method with brutal force \citep{PeFu17}. Though various sequential allocation algorithms have been proposed, there is no previous work that applies them to material selection in ALT, where usually we encounter censored observations from experiments, as discussed later in Section \ref{sec:des}. To overcome the inconvenience brought by the incomplete information, our work builds an approximate Bayesian model to learn the reliability performance of the materials, which allows us to apply the sequential allocation algorithms more efficiently in ALT.

\subsection{Overview}

The rest of the article is organized as follows. 
Section \ref{sec:des} provides detail description of our problem.  
Section \ref{sec:model} investigates the approximate Bayesian inference approach for the log-normal model and its corresponding theoretical properties. Section \ref{sec:design} develops the design criterion for sequential selection. Section \ref{sec:num} compares  the proposed approach with other test planning approaches using numerical examples. Section \ref{sec:con}
concludes the paper with discussion and future directions.

\section{Problem Description}\label{sec:des}

ALT mostly considers  different levels of the stress factors in testing and validating the reliability performance of a given product or material, which is often characterized by a lifetime model. In our problem, both stress factors and material features of the product are included in
the test planning stage. The stress factors are denoted by a $d$ dimensional vector $\bm v$, whereas the material features are denoted by a $p$ dimensional vector $\bm z$.
The stress factors are usually numerical variables providing the accelerated stress levels, such as temperature and humidity. The entries of the material feature vector $\bm z$ can be continuous variables indicating the key metrics of material characteristics, and they can also be categorical variables referring to different material types. For example, the material features may include the composition percentage of different elements in an alloy,  as well as different types of metallurgical procedures (e.g., annealing, tempering, electroplating, etc.) used to process materials.

We assume that the mean performance of material reliability can be expressed by $\mu(\bm z, \bm v; \bm \beta)$
as a function of stress factors $\bm v$ and material features $\bm z$ with an unknown parameter vector $\bm\beta$. A higher value of $\mu(\bm z, \bm v; \bm \beta)$ indicates that the corresponding 
material setting $\bm z$ leads longer material lifetime in average under the stress level combination $\bm v$. Therefore, the goal of our problem is to find the material alternative $\bm z$ which leads the best mean reliability performance under the target stress levels $\bm v^\ast$:
\begin{equation}\label{eq:opt}
\bm z^\ast(\bm v^\ast)\in\mathrm{argmax}_{\bm z\in\mathcal Z}\mu(\bm z, \bm v^\ast; \bm \beta),
\end{equation}
where $\mathcal Z$ is a set of candidate material settings in our experiments.

Since the testing process (e.g., the material wear process as in Section \ref{sec:case}) can be extremely complex, it is almost impossible to develop an accurate mathematical model for the mean material lifetime
under multiple stress factors and material features. To solve this problem, a log-normal model is often used to surrogate the material lifetime \citep{meeker2014statistical}:
\begin{equation}\label{eq:lognormal}
    \log(T)=\bm x (\bm z, \bm v)^\top\pmb\beta+\varepsilon,
\end{equation}
where $T$ is a random variable representing the lifetime of a test unit with experimental setting $\bm x(\bm z, \bm v)$, $\varepsilon$ is the error term following a normal distribution with mean zero and variance $\sigma^2$, and $\bm x(\bm z, \bm v)$ collects the intercept,  the stress factors $\bm v$, the material features $\bm z$, and the interactions between material features and stress factors. In particular, 
\begin{equation}\label{eq:x}
    \bm x(\bm z, \bm v)=\left(1, \bm v^\top, \bm z^\top, (\bm z\otimes\bm v)^\top\right)^\top,
\end{equation}
where $\bm z\otimes\bm v$ denotes the Kronecker product of $\bm z$ and $\bm v$, which is a $d\times p$ dimensional vector representing the interaction between material features and stress factors.  To simplify the notation, we reduce $\bm x(\bm z, \bm v)$ to $\bm x$ when there is no confusion. 
The linear coefficient $\pmb\beta$ is a $(p+1)\times (d+1)$ dimension vector. 
After collecting life times $T_i$'s from test units $i=1, \ldots, N$, the model parameters can be estimated via the maximum likelihood method. 

In reliability studies, the lifetime $T_i$'s are often given as the censored 
observations. Even under accelerated stress levels, the lifetime of a test unit can be as long as weeks or months. Thus, in the experimental stage,  the tests will be terminated after a given observation time $\tau_i$, even if the failure has not been observed. 
In additional to $T_i$, the failure of the $i$-th test is often recorded by a binary variable $\delta_i$. If $\delta_i=1$, failure is observed, and $T_i$ is the lifetime of the $i$-th test unit. If $\delta_i=0$, we only know that the lifetime $T_i$ is greater than $\tau_i$. Under the assumption of the log-normal model in \eqref{eq:lognormal}, the likelihood function of $\pmb\beta$ and $\sigma^2$ is
\[
L(\pmb\beta, \sigma^2|\{T_i, \tau_i, \delta_i,\bm x_i\}^N_{i=1})=\prod^N_{i=1}\left\{\frac{1}{\sigma T_i}\phi\left(\frac{\log T_i-\bm x^\top_i\pmb\beta}{\sigma}\right)\right\}^{\delta_i}
\]
\begin{equation}\label{eq:mle}
\cdot\left\{1-\Phi\left(\frac{\log \tau_i-\bm x^\top_i\pmb\beta}{\sigma}\right)\right\}^{1-\delta_i},
\end{equation}
where $\phi$ and $\Phi$ are the probability density function and the cumulative distribution function of the standard normal random variable,
respectively.

Under the linear model setting, it is critically important to develop efficient experimental design approach to solve the optimization problem in \eqref{eq:opt}. Since our goal is to find the optimal material setting more efficiently, we develop 
 a sequential optimal learning framework for ALT. Without loss of generality, we assume that the test lab 
 is only equipped with one set of test machine. Thus, 
 in each step of this sequential procedure, we only select one design point and allocate it to one test unit. The collected reliability information is used to update our belief regarding to the mean lifetime, and our belief regarding to the mean reliability performance of different material settings is used to determine the design for the next test unit.  There are two main 
 tasks under this development: 1) how to update the beliefs regarding the mean reliability performance of different material settings under the linear model setting with censored observations;
 2) how to develop experimental design criterion to select new design points at each step. In this paper, we first develop the updating formula for our belief of the mean lifetime in Section \ref{sec:model}, and then develop a policy to allocate experimental setting based on the updated belief in Section \ref{sec:design}.

\section{Approximate Bayesian Inference for Log-normal Model with Incomplete Observations}\label{sec:model}


In this section, we develop Bayesian update formulas for the log-normal model in \eqref{eq:lognormal}. 
Under the linear model setting in \eqref{eq:lognormal}, we assume that the prior of the linear coefficients $\pmb \beta$ is a multivariate normal distribution with mean $\pmb{\theta}_0$ and variance matrix $\Sigma_0$. If the lifetime $T_i$ is not censored, the conjugacy property of the multivariate normal distribution also leads to a multivariate normal posterior distribution of 
$\pmb\beta$.  For $n=1, \ldots, N$, we denote $\pmb\theta_{n}$ and $\Sigma_{n}$ as the mean vector and variance matrix of the posterior distribution of $\pmb\beta$ after including 
observations from the first $n$ test units. It is straightforward to derive that
\begin{equation}\label{eq:mean_complete}
\pmb\theta_{n+1}=\pmb\theta_n+\frac{y_{n+1}-\bm x^\top_{n+1}\pmb\theta_{n}}{\sigma^2+\bm x^\top_{n+1}\Sigma_n\bm x_{n+1}}\Sigma_n\bm x_{n+1}
\end{equation}
and 
\begin{equation}\label{eq:var_complete}
\Sigma_{n+1}=\Sigma_{n}-\frac{\Sigma_n\bm x_{n+1}\bm x^\top_{n+1}\Sigma_n}{\sigma^2+\bm x^\top_{n+1}\Sigma_n\bm x_{n+1}}.
\end{equation}
where $\bm x_{n+1}$ is the design point of the $(n+1)$-st test unit, 
$y_{n+1}=\log T_{n+1}$ is the logarithm lifetime observation, and
$\sigma^2$ is the variance of the error term in \eqref{eq:lognormal}. In our development, we assume that $\sigma^2$ is known for notational convenience.

Notice that, the conjugacy property gives closed-form parameter update formulas, which further enables convenience in the development of sequential experimental policies. See for examples in \cite{frazier2008knowledge} and \cite{frazier2009knowledge}.
However, the conjugacy property does not hold if we have censored responses. An alternative method of constructing closed-form parameter update formulas under this situation is the moment-matching based
approximate Bayesian inference. This method has been used to develop Bayesian ranking and selection approaches under a multivariate normal setting in \cite{zhang2017moment}, and its statistical consistency has recently been investigated by \cite{ChRy19}.
For our problem, the idea of approximate Bayesian inference is to approximate the posterior distribution of $\pmb\beta$ as a multivariate normal distribution with mean $\pmb\theta_{n+1}$ and variance 
$\Sigma_{n+1}$, which are the first and second moments of the posterior distribution of $\pmb\beta$ given that $\delta_{n+1}=0$, i.e., $y_{n+1}>\log\tau_{n+1}$. The approximate Bayesian update formula is given in Proposition \ref{prop1}.

\begin{proposition}\label{prop1}
Assume that, at the $(n+1)$-st step, we observe $\delta_{n+1}=0$ and $y_{n+1}>\log\tau_{n+1}$. 
Under the log-normal model, and the multivariate normal prior $\pmb\beta\sim \mathrm{MVN}(\pmb{\theta}_n,
\Sigma_{n})$, the approximation Bayesian inference gives closed-form update formulas:
\begin{equation}\label{eq: mean}
   \pmb\theta_{n+1} =\pmb\theta_n+\frac{\phi(\eta_{n})}{(1-\Phi(\eta_{n}))\sqrt{\sigma^2+\bm x^\top_{n+1}\Sigma_n\bm x_{n+1}}}\Sigma_n\bm x_{n+1},
\end{equation}
and
\begin{equation}\label{eq: var}
   \Sigma_{n+1} =\Sigma_{n}-\frac{\Sigma_n\bm x_{n+1}\bm x^\top_{n+1}\Sigma_n}{\sigma^2+\bm x^\top_{n+1}\Sigma_n\bm x_{n+1}}+\frac{\Sigma_n\bm x_{n+1}\bm x^\top_{n+1}\Sigma_n}{\sigma^2+\bm x^\top_{n+1}\Sigma_n\bm x_{n+1}}
\left(1-\eta\frac{\phi(\eta_n)}{\Phi(\eta_n)}-\frac{\phi(\eta_n)^2}{\Phi(\eta_n)^2}\right)^2,
\end{equation}
where 
\begin{equation}\label{eq:eta}
\eta_n=\frac{\log\tau_{n+1}-\bm x^\top_{n+1}\pmb\theta_{n}}{\sqrt{\sigma^2+\bm x^\top_{n+1}\Sigma_n\bm x_{n+1}}},
\end{equation}
and $\pmb\theta_{n+1}$ and $\Sigma_{n+1}$ are the first and second moments of the posterior distribution of $\pmb\beta$ given that $\delta_{n+1}=0$.
\end{proposition}

If material failure is observed, \eqref{eq:var_complete} indicates that the variance reduction is $\frac{\Sigma_n\bm x_{n+1}\bm x^\top_{n+1}\Sigma_n}{\sigma^2+\bm x^\top_{n+1}\Sigma_n\bm x_{n+1}}$. Also, if there is a censored response, the amount of variance reduction will be reduced by $\frac{\Sigma_n\bm x_{n+1}\bm x^\top_{n+1}\Sigma_n}{\sigma^2+\bm x^\top_{n+1}\Sigma_n\bm x_{n+1}}(1-\eta_n\frac{\phi(\eta_n)}{\Phi(\eta_n)}-\frac{\phi(\eta_n)^2}{\Phi(\eta_n)^2})^2$ as in \eqref{eq: var}. 
However, in sequential update, the effects of this additional term to the variance reduction
 is usually negligible. This is because that the variance $\Sigma_n$
is small when $n$ is large enough. Our numerical results often show that the variance update formulas in \eqref{eq:var_complete} and \eqref{eq: var} lead to approximately equal variances. Therefore, in terms of the variance update, we adopt \eqref{eq:var_complete} for both complete and censored responses. As a result, the update formulation at step $n$ can be summarized by
\begin{eqnarray}
\pmb\theta_{n+1} &=& \pmb\theta_n + \delta_{n+1}\frac{y_{n+1}-\bm x^\top_{n+1}\pmb\theta_{n}}{\sigma^2+\bm x^\top_{n+1}\Sigma_n\bm x_{n+1}}\Sigma_n\bm x_{n+1}
 \nonumber\\
&& + (1 -  \delta_{n+1}) \frac{\phi(\eta_{n})}{(1-\Phi(\eta_{n}))\sqrt{\sigma^2+\bm x^\top_{n+1}\Sigma_n\bm x_{n+1}}}\Sigma_n\bm x_{n+1},
, \nonumber\\
\Sigma_{n+1}&=&\Sigma_{n}-\frac{\Sigma_n\bm x_{n+1}\bm x^\top_{n+1}\Sigma_n}{\sigma^2+\bm x^\top_{n+1}\Sigma_n\bm x_{n+1}},
\label{eq:sum_update}
\end{eqnarray}
with $\eta_n$ given in \eqref{eq:eta}.


We now discuss the consistency property of the proposed approximate Bayesian inference under incomplete observations. 
In the following context, we demonstrate the convergence of the sequence $\left(\pmb\theta_n\right)_{n=0}^\infty$ based on the framework established in \cite{ChRy19}. We make the following assumptions:
\begin{assumption} \label{chena1}
The design vectors $\left(\bm x_n\right)_{n=0}^\infty$ are drawn i.i.d. from a common distribution satisfying $\mathrm{E} \left(\bm x_n \bm x_n^\top\right) = \mathbf{A}$, where $\mathbf{A}$ is a positive definite symmetric matrix.
\end{assumption}
\begin{assumption} \label{chena2}
The sequence $\left(\bm x_n\right)_{n=0}^\infty$ satisfies $0 < \inf_n ||\bm x_n||_1 \leq \sup_n ||\bm x_n||_1 < \infty$ almost surely. 
\end{assumption}

\begin{theorem} \label{chent1}
Suppose Assumptions \ref{chena1}-\ref{chena2} hold and the sequence $\left(\log \tau_n\right)_{n=0}^\infty$ is bounded, and suppose that $\pmb\theta_n$ and $\Sigma_n$ are updated using (\ref{cheneq1})-(\ref{cheneq2}) respectively. Then, $\pmb\theta_n \to \pmb\beta$ almost surely.
\end{theorem}

The proof of this Theorem is deferred to the Appendix. This theorem indicates that although we approximate the posterior distribution 
to a multivariate normal under censored observations, the approximation can be asymptotically accurate, since the updated parameter sequence $\left(\pmb\theta_n\right)_{n=0}^\infty$ converges to the true model parameters.

\section{Sequential Selection for Reliability Improvement}\label{sec:design}

This section discusses how to select design points in a sequential manner. As mentioned earlier, we investigate a fully sequential procedure, and assume that only one experimental unit will be allocated in each step of the sequential procedure. Recall that our goal is to determine the material feature combination $\bm z^\ast(\bm v^\ast)$ such that it has the best reliability performance under the target stress factor levels  $\bm v^\ast$.
At the $n$-th step of the sequential procedure, the optimal material setting based on the collected information can be expressed by
\begin{equation}\label{eq:opt_n}
\bm z^n(\bm v^\ast)\in\mathrm{argmax}_{\bm z\in\mathcal Z}\mathrm{E}^n\mu(\bm z, \bm v^\ast;\pmb\beta),
\end{equation}
where $\mathrm{E}^n$ represents that the expectation is taken with respect to the prior distribution of $\pmb\beta$ at the $n$-th step.
Under the log-normal model setting in \eqref{eq:lognormal}, the objective in \eqref{eq:opt_n} can be simplified to 
\[
\mathrm{E}^n\mu(\bm z, \bm v^\ast;\pmb\beta)=\mathrm{E}^n \left[\bm x(\bm z, \bm v^\ast)^\top\pmb\beta\right]=\bm x(\bm z, \bm v^\ast)^\top\pmb\theta_n
\]
with $\bm x(\bm z, \bm v^\ast)$ given in \eqref{eq:x}. To meet the requirement of our goal in \eqref{eq:opt}, 
new design points in each step should be determined to maximize the improvement the target optimization problem. The improvement of the objective in \eqref{eq:opt} by adding new design points in the $(n+1)$-st step can be quantified by
\[
\mathrm{max}_{\bm z\in\mathcal Z}\mathrm{E}^{n+1}\mu(\bm z, \bm v^\ast; \pmb\beta)-\mathrm{max}_{\bm z\in\mathcal Z}\mathrm{E}^n\mu(\bm z, \bm v^\ast; \pmb\beta)
\]
\begin{equation}\label{eq:improve}
=\mathrm{max}_{\bm z\in\mathcal Z}\left[\bm x(\bm z, \bm v^\ast)^\top\pmb\theta_{n+1}\right]-\mathrm{max}_{\bm z\in\mathcal Z}\left[\bm x(\bm z, \bm v^\ast)^\top\pmb\theta_n\right].
\end{equation}
Since $\pmb\theta_{n+1}$ is a random vector that depends on the selected design points $\bm x_{n+1}=\bm x(\bm z_{n+1}, \bm v_{n+1})$, the $(n+1)$-st design point should be chosen to maximize the expectation of the value of improvement given that $(\bm z, \bm v)$ is the design point at the $(n+1)$-st step. Therefore, the acquisition function to select the new design point can be expressed by
\[
\mathrm{EI}^n(\bm z, \bm v; \bm v^\ast)=\mathrm{E}\left\{\mathrm{max}_{\bm z'\in\mathcal Z}\left[\bm x(\bm z', \bm v^\ast)^\top\pmb\theta_{n+1}\right]|\bm z_{n+1}=\bm z, \bm v_{n+1}=\bm v\right\}
\]
\begin{equation}\label{eq:kg}
-\mathrm{max}_{\bm z'\in\mathcal Z}\left[\bm x(\bm z', \bm v^\ast)^\top\pmb\theta_n\right],
\end{equation}
where the expectation is taken with respect to the posterior predictive distribution of $y_{n+1}$ given that $\bm z_{n+1}=\bm z$ and $\bm v_{n+1}=\bm v$ are the $(n+1)$-st design point. This EI-type acquisition function is typically used in selecting design points for optimization problem in a sequential manner, see \cite{powell2012optimal} for examples of the EI-type acquisition function under different developments.

For our problem, \eqref{eq:kg} can be further simplified. 
Since $\pmb\theta_{n+1}$ with non-censored response
is given by \eqref{eq:mean_complete},  we have that
\[
\bm x(\bm z, \bm v^\ast)^\top\pmb\theta_{n+1}=\bm x(\bm z, \bm v^\ast)^\top\pmb\theta_{n}+\frac{y_{n+1}-\bm x^\top_{n+1}\pmb\theta_{n}}{\sigma^2+\bm x^\top_{n+1}\Sigma_n\bm x_{n+1}}\bm x(\bm z, \bm v^\ast)^\top\Sigma_n\bm x_{n+1}.
\]
Under the log-normal model and the prior distribution of $\pmb\beta\sim\mathrm{MVN}(\pmb\theta_{n}, \Sigma_{n})$, it is straightforward to derive that the posterior predictive distribution of $y_{n+1}$ is a normal distribution with mean $\bm x^\top_{n+1}\pmb\theta_{n}$ and variance $\sigma^2+\bm x^\top_{n+1}\Sigma_n\bm x_{n+1}$. Therefore, we can express
\begin{equation}\label{eq:pred_theta}
\bm x(\bm z, \bm v^\ast)^\top\pmb\theta_{n+1}=\bm x(\bm z, \bm v^\ast)^\top\pmb\theta_{n}+\frac{\bm x(\bm z, \bm v^\ast)^\top\Sigma_n\bm x_{n+1}}{\sqrt{\sigma^2+\bm x^\top_{n+1}\Sigma_n\bm x_{n+1}}}G,
\end{equation}
where $G$ is a standard normal random variable.

We denote 
$\tilde {\bm v}^\ast=(1,  (\bm v^\ast)^\top)^\top$. Then $\bm x(\bm z, \bm v^\ast)^\top\pmb\theta_n=(\tilde{\bm v}^\ast)^\top\pmb\theta_{n, 0}+(\bm z\otimes\tilde{\bm v}^\ast)^\top\pmb\theta_{n,1}$, 
where $\pmb\theta_n=(\pmb\theta^\top_{n,0}, \pmb\theta^\top_{n,1})^\top$ with $\pmb\theta_{n,0}$ and $\pmb\theta_{n,1}$ being 
vectors of size $d+1$ and $p(d+1)$, respectively. Accordingly, 
\begin{equation}\label{eq:theta_n}
 \mathrm{max}_{\bm z\in\mathcal Z}\left[\bm x(\bm z, \bm v^\ast)^\top\pmb\theta_n\right]=(\tilde{\bm v}^\ast)^\top\pmb\theta_{n, 0}+\mathrm{max}_{\bm z\in\mathcal Z}\left[(\bm z\otimes\tilde{\bm v}^\ast)^\top\pmb\theta_{n,1}\right],
\end{equation}
and
\[
\mathrm{max}_{\bm z\in\mathcal Z}\left[\bm x(\bm z, \bm v^\ast)^\top\pmb\theta_{n+1}\right]=\mathrm{max}_{\bm z\in\mathcal Z}\left\{\bm x(\bm z, \bm v^\ast)^\top\pmb\theta_{n}+\frac{\bm x(\bm z, \bm v^\ast)^\top\Sigma_n\bm x_{n+1}}{\sqrt{\sigma^2+\bm x^\top_{n+1}\Sigma_n\bm x_{n+1}}}G\right\}
\]
\begin{equation}\label{eq:maxmodel}
=(\tilde{\bm v}^\ast)^\top\pmb\theta_{n, 0}+\mathrm{max}_{\bm z\in\mathcal Z}\left\{(\bm z\otimes\tilde{\bm v}^\ast)^\top\pmb\theta_{n,1}+\frac{\bm x(\bm z, \bm v^\ast)^\top\Sigma_n\bm x_{n+1}}{\sqrt{\sigma^2+\bm x^\top_{n+1}\Sigma_n\bm x_{n+1}}}G\right\}.
\end{equation}

Plugging \eqref{eq:theta_n} and \eqref{eq:maxmodel} into \eqref{eq:kg}, we obtain that
\[
\mathrm{EI}^n(\bm z, \bm v; \bm v^\ast)=\mathrm{E}_G\left\{\mathrm{max}_{\bm z'\in\mathcal Z}\left[(\bm z'\otimes\tilde{\bm v}^\ast)^\top\pmb\theta_{n,1}+\frac{\bm x(\bm z', \bm v^\ast)^\top\Sigma_n\bm x(\bm z, \bm v)}{\sqrt{\sigma^2+\bm x^\top(\bm z, \bm v)\Sigma_n\bm x(\bm z, \bm v)}}G\right]\right\}
\]
\begin{equation}\label{eq: reducedEI}
-\mathrm{max}_{\bm z'\in\mathcal Z}\left[(\bm z'\otimes\tilde{\bm v}^\ast)^\top\pmb\theta_{n,1}\right],
\end{equation}
where the expectation $\mathrm{E}_G$ is taken with respect to the random variable $G$. The new design point $\bm x_{n+1}=\bm x(\bm z_{n+1}, \bm v_{n+1})$ is selected to maximize this acquisition function. 

For our problem, the number of candidate material settings in $\mathcal Z$ is often finite, say, $\mathcal Z=\{\bm z^1, \ldots, \bm z^K\}$. Under this situation, 
$\mathrm{EI}^n(\bm z, \bm v; \bm v^\ast)$ has a closed-form expression according to \cite{frazier2009knowledge}. Let 
\[
b^k_n(\bm z, \bm v; \bm v^\ast)=\frac{\bm x(\bm z^k, \bm v^\ast)^\top\Sigma_n\bm x(\bm z, \bm v)}{\sqrt{\sigma^2+\bm x^\top(\bm z, \bm v)\Sigma_n\bm x(\bm z, \bm v)}}
\]
for $k=1, \ldots, K$. For notational convenience, we assume that $b^k_n(\bm z, \bm v; \bm v^\ast)<b^{k+1}_n(\bm z, \bm v; \bm v^\ast)$
for $k=1,\ldots, K-1$. Following \cite{frazier2009knowledge}, we have that 
\[
\mathrm{EI}^n(\bm z, \bm v; \bm v^\ast)=\sum^K_{k=1} \left[b^{k+1}_n(\bm z, \bm v; \bm v^\ast)-b^k_n(\bm z, \bm v; \bm v^\ast)\right]
\]
\begin{equation}\label{eq:closedEI}
\cdot g\left\{-\frac{|(\bm z^{k+1}\otimes\tilde{\bm v}^\ast)^\top\pmb\theta_{n,1}-(\bm z^k\otimes\tilde{\bm v}^\ast)^\top\pmb\theta_{n,1}|}{b^{k+1}_n(\bm z, \bm v; \bm v^\ast)-b^k_n(\bm z, \bm v; \bm v^\ast)}\right\},
\end{equation}
where $g(u)=u\Phi(u)+\phi(u)$.
To maximize $\mathrm{EI}^n(\bm z, \bm v; \bm v^\ast)$, we can compute its gradient with regard to $\bm v$ according to \cite{zhang2019}, and use gradient based optimization approaches to find the maximum of  $\mathrm{EI}^n(\bm z, \bm v; \bm v^\ast)$ for each given $\bm z\in\mathcal Z$. 

Notice that, the EI-type sequential design criterion in \eqref{eq:kg} may not lead to a closed-form expression as in \eqref{eq:closedEI} if the posterior of the coefficients $\pmb\beta$ is not a multivariate normal distribution in each step. The proposed approximation Bayesian update in Section \ref{sec:model} guarantees that 
the multivariate normal posterior distribution holds. Besides convenient and efficient model update, the proposed Bayesian approximation also plays an important role in simplifying the computation of sequential design selection.

\section{Numerical Study}\label{sec:num}

This section provides synthetic examples and a case study on accelerated wear testing to compare the numerical performances of 
different model updates and experimental design approaches. In terms of model updates, we compare the proposed approximation Bayesian update formulas in \eqref{eq:sum_update}  
with the exact update, i.e., refitting the log-normal model using all the data points, which does not possess tractable parameter updating formulas.  
Those two alternatives approaches are denoted by ``approx'' and ``exact'', respectively. 
We also consider the following experimental design approaches:
\begin{itemize}
    \item[1.] ({\bf Design:}) Full factorial designs, see for example, \cite{wu2011experiments}.
    \item[2.] ({\bf SeqD:}) Sequential Bayesian D-optimal Design in \cite{lee2018sequential}.
    \item[3.] ({\bf SeqEI:}) The EI-based sequential design procedure described in Section  \ref{sec:design}.
\end{itemize}
We consider all possible combinations of the two model update approaches and the three experimental design approaches. The six alternatives involved in our numerical comparison are denoted by ``Design approx'',  ``Design exact'',  ``SeqD approx'',  ``SeqD exact'',  ``SeqEI approx'',  and ``SeqEI exact'', respectively. 

Notice that, the EI-type sequential design criterion in \eqref{eq:kg} may not lead to a closed-form expression as in \eqref{eq:closedEI} if the posterior of the coefficients $\pmb\beta$ is not a multivariate normal distribution. 
For ``SeqEI approx'', our model (the posterior distribution of $\pmb\beta$) can be represented by a multivariate normal distribution completely, based on the proposed approximation Bayesian update in Section \ref{sec:model}. Thus, the proposed Bayesian approximation also plays an important role in simplifying the computation of sequential design selection. 
However, under the exact model update, the implementation of this EI-type sequential design criterion is impractical, since it may require MCMC to approximate the value of \eqref{eq:kg} for each candidate $\bm v$ and $\bm z$ at each step. 
In our implementation of ``SeqEI exact'', we process the model and the experimental design selection under two separate tracks: the design criterion in \eqref{eq:closedEI} is obtained under the proposed approximate model update (the same as in ``SeqEI approx''), whereas the collected data points are used to refit the exact model and determine the optimal material setting according to \eqref{eq:opt_n} at each step.
In this way, we can evaluate the effects of model update and sequential design separately.

The full factorial designs are one-shot designs, which are not originally developed for a sequential experimentation. To compare the full factorial design under a sequential manner, we make it adaptable for a sequential procedure.  First, we generate a full factorial design with respect to the number of levels of the material feature factors and the stress factors. Since the total number of steps $N$ is usually greater than the run size of this full factorial design, we replicate the runs in the full factorial design one by one to make total run size equal to $N$ (i.e., the runs in original full factorial design may not have exact equal number of replications). Finally, we randomize the order of the design within the $N$ runs, and let them enter the sequential procedure one by one.

The goal of our problem is to choose the material setting with the best reliability performance. In practice, we often consider a finite number of material settings. Thus, we consider discrete levels of the material factors, and use probability of correct selection at the target stress level $\bm v^\ast$ to evaluate different approaches. 
According to \eqref{eq:opt} and \eqref{eq:opt_n}, the probability of correct selection 
can be expressed by $\mathrm{P}(\bm z^n(\bm v^\ast)=\bm z^\ast(\bm v^\ast))$, where the probability is taken with respect to $\bm z^n(\bm v^\ast)$, which is a 
random variable due to the randomness of collected responses. 
In our numerical study, the probability of correct selection is estimated empirically by
\begin{equation}\label{eq:eps}
    \hat{\mathrm{P}}(\bm z^n(\bm v^\ast)=\bm z^\ast(\bm v^\ast))=\frac{1}{R}\sum^R_{r=1}I(\bm z^n_r(\bm v^\ast)=\bm z^\ast(\bm v^\ast)),
\end{equation}
where $R$ is the total number of replications, $I(\cdot)$ is an indicator function, and $\bm z^n_r(\bm v^\ast)$ is the selected optimal material setting at the $n$-th step from the $r$-th replication. In the synthetic examples and the case study, we use $R=100$ to compute the estimated probability of correct selection.
In all of our numerical examples, we set the observation time $\tau_i$ in \eqref{eq:mle}
to be a constant.

\subsection{Synthetic Examples}\label{sec:syn}

In this study, we directly generate data from the log-normal model in \eqref{eq:lognormal}.
The stress factor $\bm v$ contains three dimensions. For each dimension, the design points of the accelerated lab experiments are taken value from $\{0.5, 1\}$, whereas the targeted environmental condition is specified to be 0.1. For the material factors, we generate one factor with $K$ levels. The first level of this material factor is specified to be optimal with the best reliability performance in average.  
We generate four random variables from uniform distribution $U(-1/30, 0)$ to be 
the linear coefficients corresponding to the intercept and each of three stress factors. The generated four dimensional linear coefficients are denoted by a vector
 $\bm \beta_1$.  The linear coefficients of each remaining material level are generated by $\bm \beta_1+\bm \beta_k$ for $k=2, \ldots, K$, where each component of $\bm \beta_k$ is a uniform random variable from -1/30 to 0. This setting guarantees that the first level of the material factor has the best reliability performance in average, and the average lifetime decreases as stress factor levels increase. A total number of 100 replications is used to estimate the probability of correct selection as in \eqref{eq:eps}. For each replication, we generate 20 data points for each material setting to obtain the prior distributions for the linear coefficients.

\begin{figure}[!h]
    \centering
    \includegraphics[scale=0.65]{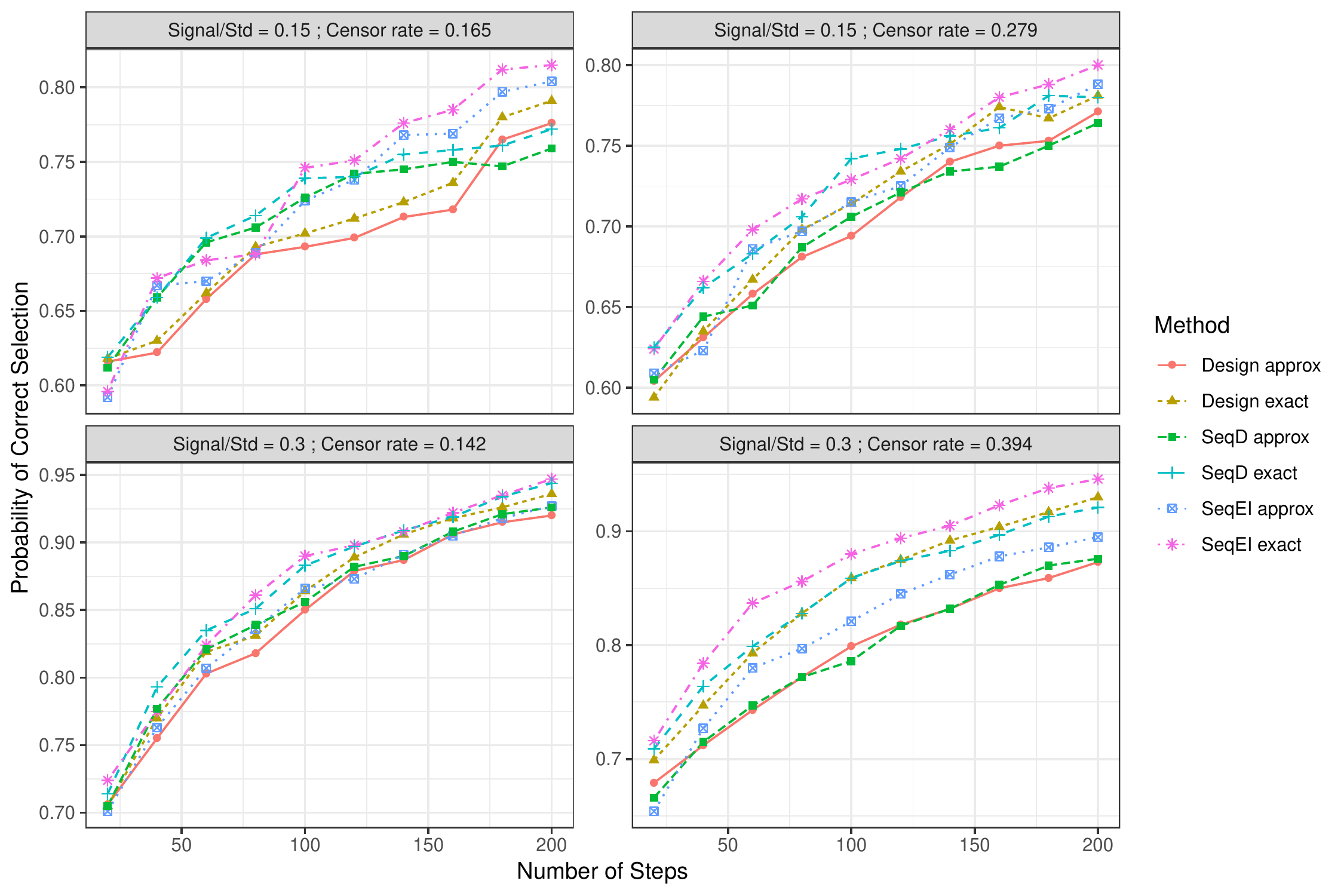}
    \caption{The estimated probability of correct selection for different settings with
    $K=2$.}
    \label{fig:ex_level2}
\end{figure}

\begin{figure}[!h]
    \centering
    \includegraphics[scale=0.65]{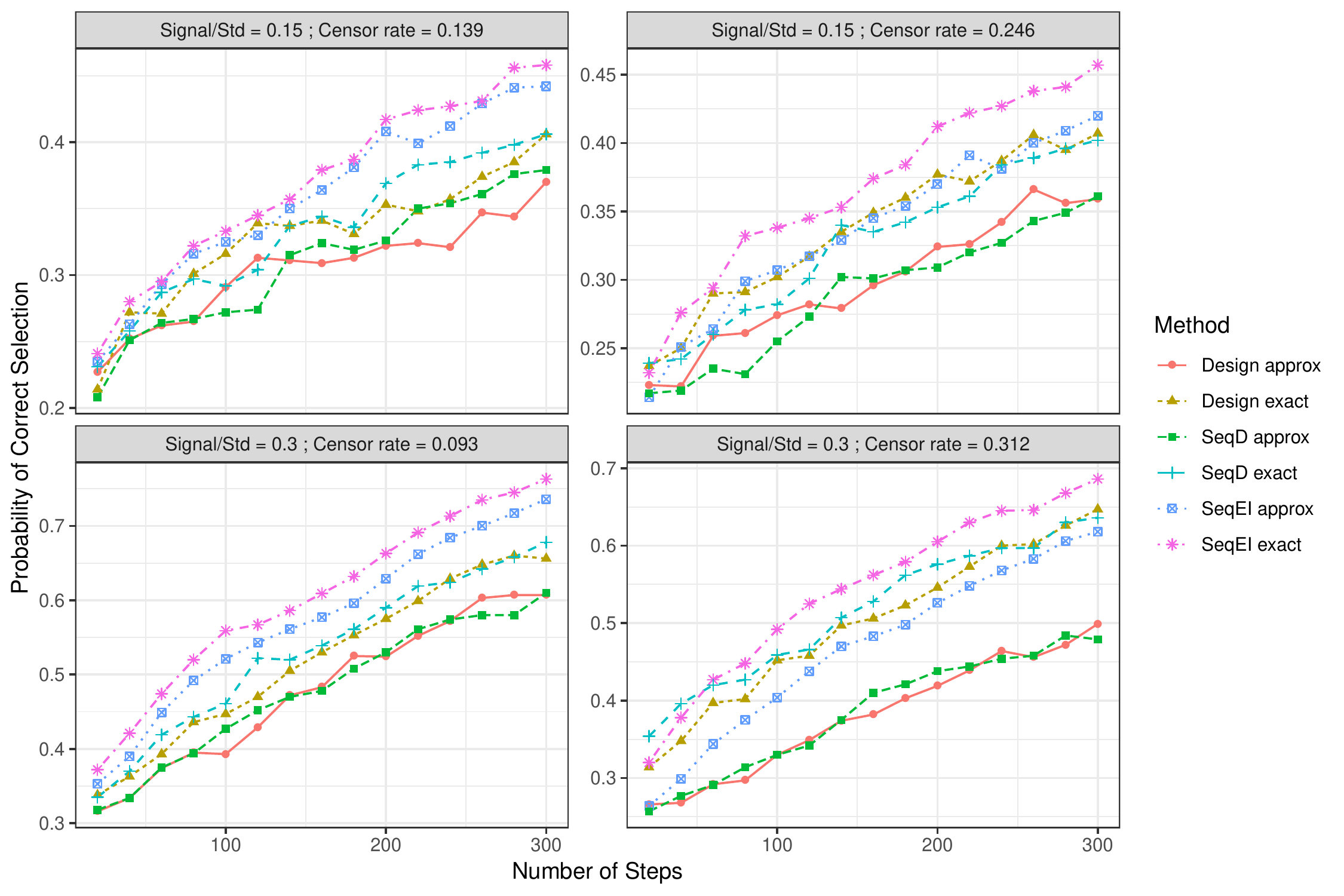}
    \caption{The estimated probability of correct selection for different settings with $K=6$.}
    \label{fig:ex_level6}
\end{figure}

In Figure \ref{fig:ex_level2}, we consider a case with only two material settings, i.e.,  $K=2$. We generate the responses under different signal to noise ratios. The signal level (i.e., the value of coefficients) is fixed as described earlier. The value of standard deviation $\sigma$ in \eqref{eq:lognormal} is set to be 0.2 or 0.1, and resulted value of ``Signal/Std'' is 0.15
as in the top panel of Figure \ref{fig:ex_level2} or 0.3 as in the bottom panel of Figure \ref{fig:ex_level2}. 
The value of the constant observational time $\tau_i$ in \eqref{eq:mle} is
set to be 1 or 1.2 to generate different levels of response censor rates. As shown in Figure \ref{fig:ex_level2}, 
the censoring rate is around 15\% if $\tau_i=1.2$ (left panel), whereas the censoring rate is around
30\% if $\tau_i=1$ (right panel). Under a similar setting, we show the results of a scenario with 
six material settings (i.e., $K=6$) in Figure \ref{fig:ex_level6}.

The results in Figures \ref{fig:ex_level2}-\ref{fig:ex_level6} show that ``SeqEI'' based approaches give the highest probability of correct selection. Since the design criterion  of ``SeqEI'' is developed to improve the optimization problem in \eqref{eq:opt}, it outperforms ``Design'' and ``SeqD'', both of which aim for reducing the variances of model coefficients. 
We also see that, ``approx'' approach does not perform well if the censoring rate is high (say, around 30\%). It demonstrates that the efficiency of the proposed approximate model updating approach can deteriorate if there is a significant large portion of censored observations. Overall, ``SeqEI exact'' gives the best performance, and the performance of ``SeqEI approx'' is competitive to the best when the censoring rate is low. 
For challenging scenarios (e.g., ``Signal/Std=0.15'' or $K=6$), ``SeqEI'' based approaches demonstrate obvious advantages compared to other design approaches.

\subsection{A Case Study on Accelerated Wear Tests}\label{sec:case}


We consider a material wear test of copper alloys as an example to demonstrate the performance of the proposed sequential selection method. Because of high strength and exceptional bearing properties of copper alloys, they are widely considered in various safety-/mission-critical industries, e.g., aircraft bearings and bushings in aerospace industry, drilling and mining equipment in mining industry. This case study considers the reliability performance of Cu-Ni-Sn alloys in the accelerated wear tests. This study investigates two types of material specimens, namely as-received Cu-Ni-Sn and annealed Cu-Ni-Sn specimens. Due to the annealing process, the microstructures as well as physical/chemical properties of annealed Cu-Ni-Sn specimens will be altered as compared to the as-received ones. Thus, their reliability performances may differ accordingly. The experimenter is interested in finding the material with better reliability performance. Wear tests were carried out using a Koehler K93500 pin-on-disc tester under various environmental conditions of ``Load'', ``Temperature'' and ``Humidity'' \citep{singh2007dry}. For each testing unit of Cu alloy specimens, in-situ monitoring outputs of wear performance (e.g., wear depth in $\mu$m) are measured over time by a linear variable displacement transducer. A material failure is recorded if the material weight loss is above a given threshold value.
Historical data contains the information of the wearing processes of 18 experimental units. 

The experimental observations of all 18 experimental units are provided for our study. Unfortunately, follow-up experiments are not available to validate the proposed sequential design approach. Therefore, 
to implement the sequential selection procedure, we develop a pseudo simulator to model the historical data. This pseudo simulator is built on a Gaussian process model. Under this pseudo simulator, the log response is not a linear function of the material factor and stress factors. We are able to investigate the robustness of the proposed approach under this nonlinear setting. 
The goal of this case study is to choose the materiel option that maximizes the reliability performance. According to the evidence shown from the data and domain knowledge, we identify that as-received Cu-Ni-Sn alloy is more reliable than annealed Cu-Ni-Sn alloy. With this information, we are able to estimate the probability of correct selection as in \eqref{eq:eps}. In this study, we consider that the observation times $\tau_i$ equal to 200, 300, and 500 to generate different censoring rate of the responses.

\begin{figure}
    \centering
    \includegraphics[scale=0.55]{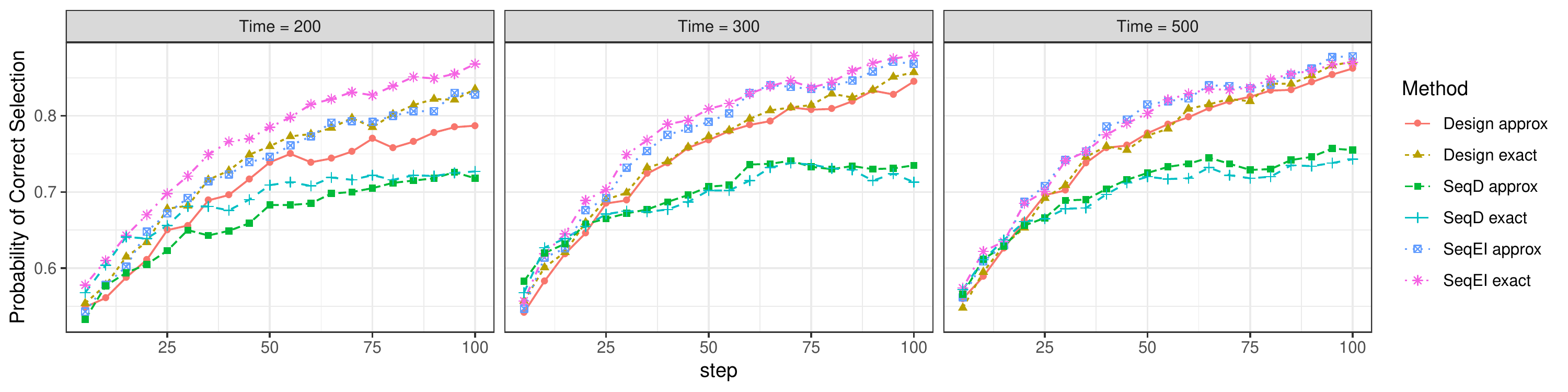}
    \caption{The estimated probability of correct selection for the case study.}
    \label{fig:ex_case}
\end{figure}

The results of different approaches are shown in Figure \ref{fig:ex_case}. The censoring rates corresponding to observational times 200, 300, and 500 are  45.1\%, 34.7\%, and 29.7\%, respectively.
Similar to the results from Section \ref{sec:syn},  ``SeqEI exact'' gives the best performance in general, and the performance of ``SeqEI approx'' is competitive to the best when the censoring rate is low.

\section{Conclusion}\label{sec:con}

This paper proposed a sequential test planning approach to  
determine the most reliable material setting in accelerated lab experiments. 
To guarantee a tractable statistical mechanism for information collection and update, we develop explicit model  parameter update formulas via approximate Bayesian inference. We demonstrate the advantage of our proposal through theoretical results and numerical studies. Now we remark on the directions for future research. First, we assume that the observation times for each experimental unit is given in this paper. It is more practical and efficient to determine the observational time for each test unit based on existing experimental results. The decision of allocating  observation time  to each test unit can be more critical when there is a deadline to complete all experiments. Second, this paper considers a single operation condition. In some other studies, the target levels of the stress factors might be different under different practical situations, each of which might be prone to different material settings. It is interesting to extend our work to this personalized optimization scheme, and develop a sequential selection procedure to choose the optimal material settings for each individualized environmental situation.

 \appendix 
\section{Proof of Proposition \ref{prop1}}

First of all, according to the assumption of the log-normal model, we have that 
\[
y_{n+1}\sim N\left(\bm x^\top_{n+1}\pmb\theta_{n}, \sigma^2+\bm x^\top_{n+1}\Sigma_n\bm x_{n+1}\right).
\]
Then $y_{n+1}|y_{n+1}>\log\tau_{n+1}$ follows a truncated normal distribution (see for an example, \cite{johnson1970continuous}), and its mean and variance are given by 
\[
\mathrm{E}(y_{n+1}|y_{n+1}>\log\tau_{n+1})=\bm x^\top_{n+1}\pmb\theta_{n}+
\frac{\phi(\eta_n)}{1-\Phi\left(\eta_n\right)}\sqrt{\sigma^2+\bm x^\top_{n+1}\Sigma_n\bm x_{n+1}}
\]
and
\[
\mathrm{Var}(y_{n+1}|y_{n+1}>\log\tau_{n+1})=(\sigma^2+\bm x^\top_{n+1}\Sigma_n\bm x_{n+1})\left(1-\eta\frac{\phi(\eta_n)}{\Phi(\eta_n)}-\frac{\phi(\eta_n)^2}{\Phi(\eta_n)^2}\right)^2.
\]
According to \eqref{eq:mean_complete} and \eqref{eq:var_complete}, we have that 
\[
\pmb\beta|y_{n+1}\sim \mathrm{MVN}\left(\pmb\theta_n+\frac{y_{n+1}-\bm x^\top_{n+1}\pmb\theta_{n}}{\sigma^2+\bm x^\top_{n+1}\Sigma_n\bm x_{n+1}}\Sigma_n\bm x_{n+1}, \Sigma_{n}-\frac{\Sigma_n\bm x_{n+1}\bm x^\top_{n+1}\Sigma_n}{\sigma^2+\bm x^\top_{n+1}\Sigma_n\bm x_{n+1}}\right),
\]
Therefore, the posterior mean and variance of $\pmb\beta$ given $y_{n+1}>\log\tau_{n+1}$ can be derived by
\[
\mathrm{E}(\pmb\beta|y_{n+1}>\log\tau_{n+1})=\mathrm{E}\left[\mathrm{E}(\pmb\beta|y_{n+1})|y_{n+1}>\log\tau_{n+1}\right]
\]
\[=\pmb\theta_n+\frac{\mathrm{E}(y_{n+1}|y_{n+1}>\log\tau_{n+1})-\bm x^\top_{n+1}\pmb\theta_{n}}{\sigma^2+\bm x^\top_{n+1}\Sigma_n\bm x_{n+1}}\Sigma_n\bm x_{n+1}
\]
\[
=\pmb\theta_n+\frac{\phi(\eta_n)}{(1-\Phi(\eta_n))\sqrt{\sigma^2+\bm x^\top_{n+1}\Sigma_n\bm x_{n+1}}}\Sigma_n\bm x_{n+1}
\]
and
\[
\mathrm{Var}(\pmb\beta|y_{n+1}>\log\tau_{n+1})=\mathrm{E}(\mathrm{Var}(\pmb\beta|y_{n+1})|y_{n+1}>\log\tau_{n+1})+\mathrm{Var}(\mathrm{E}(\pmb\beta|y_{n+1})|y_{n+1}>\log\tau_{n+1})
\]
\[
=\Sigma_{n}-\frac{\Sigma_n\bm x_{n+1}\bm x^\top_{n+1}\Sigma_n}{\sigma^2+\bm x^\top_{n+1}\Sigma_n\bm x_{n+1}}+\frac{\Sigma_n\bm x_{n+1}\bm x^\top_{n+1}\Sigma_n}{(\sigma^2+\bm x^\top_{n+1}\Sigma_n\bm x_{n+1})^2}\mathrm{Var}(y_{n+1}|y_{n+1}>\log\tau_{n+1})
\]
\[
=\Sigma_{n}-\frac{\Sigma_n\bm x_{n+1}\bm x^\top_{n+1}\Sigma_n}{\sigma^2+\bm x^\top_{n+1}\Sigma_n\bm x_{n+1}}+\frac{\Sigma_n\bm x_{n+1}\bm x^\top_{n+1}\Sigma_n}{\sigma^2+\bm x^\top_{n+1}\Sigma_n\bm x_{n+1}}
\left(1-\eta_n\frac{\phi(\eta_n)}{\Phi(\eta_n)}-\frac{\phi(\eta_n)^2}{\Phi(\eta_n)^2}\right)^2.
\]

\section{Proof of Theorem \ref{chent1}}

From law of large number, Assumptions \ref{chena1}-\ref{chena2} lead to $\lim_{n\to \infty} \frac{1}{n}\sum_{k=0}^\infty \bm x_k \bm x_k^\top = \mathbf{A}$ almost surely. Furthermore, denote $\mathbf{B} = \frac{1}{\sigma^2} \mathbf{A}$, then by Lemma EC.2 in \cite{ChRy19}, we have the following result on the convergence rate of $\Sigma_n$.
\begin{lemma} \label{chenl1}
Suppose Assumptions \ref{chena1}-\ref{chena2} hold, then, with probability 1, 
\begin{eqnarray*}
\sum_{n=1}^\infty \frac{1}{(n+1)^{\frac{3}{4}}} \left\Vert \frac{1}{n+1}\Sigma_{n+1}^{-1} - \mathbf{B}\right\Vert_2^2 &<& \infty. 
\end{eqnarray*}
\end{lemma}
This lemma will be used in the proof of Theorem \ref{chent1}.

In the remaining of this proof, we assume that a suitable set of measure 0 is discarded, so we don't have to repeat the qualification ``almost surely''. Notice that, according to the Woodbury matrix identity \citep{woodbury1950inverting}, the updating formulas in \eqref{eq:sum_update} can be expressed by
\begin{eqnarray}
\pmb\theta_{n+1} &=& \pmb\theta_n + \delta_{n+1} \frac{y_{n+1}-\bm x^\top_{n+1}\pmb\theta_n}{\sigma^2} \Sigma_{n+1}\bm x_{n+1} \nonumber\\
&& + (1 -  \delta_{n+1}) \frac{\phi(\eta_{n}) \sqrt{\sigma^2+\bm x^\top_{n+1}\Sigma_n\bm x_{n+1}}}{(1-\Phi(\eta_{n})) \sigma^2}\Sigma_{n+1}\bm x_{n+1}, \label{cheneq1}\\
\Sigma_{n+1}^{-1} &=& \Sigma_n^{-1} + \frac{1}{\sigma^2} \bm x_{n+1}\bm x^\top_{n+1}, \label{cheneq2}
\end{eqnarray}
where $\eta_n$ is expressed in \eqref{eq:eta}. The development of the proof will be based on the expressions above.

Without loss of generality, let $\pmb\beta = 0$. Denote 
\begin{eqnarray*}
\xi_n &=& \frac{\log\tau_{n+1}-\bm x^\top_{n+1}\pmb\theta_n}{\sigma},\\
Q_n &=& - \delta_{n+1} \frac{y_{n+1}-\bm x^\top_{n+1}\pmb\theta_n}{\sigma^2}\mathbf{B}^{-\frac{1}{2}}\bm x_{n+1}  - (1 -  \delta_{n+1}) \frac{\phi(\xi_{n})}{(1-\Phi(\xi_{n})) \sigma}\mathbf{B}^{-\frac{1}{2}}\bm x_{n+1},\\
b_n &=& - \delta_{n+1} \mathbf{B}^{\frac{1}{2}} \left( \frac{y_{n+1}-\bm x^\top_{n+1}\pmb\theta_n}{\sigma^2} (n+1)\Sigma_{n+1}\bm x_{n+1} - \frac{y_{n+1}-\bm x^\top_{n+1}\pmb\theta_n}{\sigma^2}\mathbf{B}^{-1}\bm x_{n+1}\right) \nonumber\\
&& - (1 -  \delta_{n+1}) \mathbf{B}^{\frac{1}{2}} \left(\frac{\phi(\eta_{n}) \sqrt{\sigma^2+\bm x^\top_{n+1}\Sigma_n\bm x_{n+1}}}{(1-\Phi(\eta_{n})) \sigma^2} (n+1) \Sigma_{n+1}\bm x_{n+1} - \frac{\phi(\xi_{n})}{(1-\Phi(\xi_{n})) \sigma}\mathbf{B}^{-1}\bm x_{n+1}\right)\\
&=& -\delta_{n+1} \mathbf{B}^{\frac{1}{2}} \frac{y_{n+1}-\bm x^\top_{n+1}\pmb\theta_n}{\sigma^2} (n+1)\Sigma_{n+1}\bm x_{n+1} \nonumber\\
&& - (1 -  \delta_{n+1}) \mathbf{B}^{\frac{1}{2}} \frac{\phi(\eta_{n}) \sqrt{\sigma^2+\bm x^\top_{n+1}\Sigma_n\bm x_{n+1}}}{(1-\Phi(\eta_{n})) \sigma^2} (n+1) \Sigma_{n+1}\bm x_{n+1} \\
&& - Q_n. 
\end{eqnarray*}
Then, (\ref{cheneq1}) is equivalent to 
\begin{eqnarray*}
\mathbf{B}^{\frac{1}{2}} \pmb\theta_{n+1} &=& \mathbf{B}^{\frac{1}{2}}\pmb\theta_n- \frac{1}{n+1}\left(Q_n + b_n\right). 
\end{eqnarray*}
Taking the $\ell^2$-norm, we have
\begin{eqnarray}
\left\Vert \mathbf{B}^{\frac{1}{2}}\pmb\theta_{n+1}\right\Vert_2^2 &=& \pmb\theta_{n+1}^\top \mathbf{B}\pmb\theta_{n+1}\nonumber\\
&=& \left\Vert \mathbf{B}^{\frac{1}{2}}\pmb\theta_n\right\Vert_2^2 + \frac{1}{(n+1)^2} \left \Vert Q_n \right\Vert_2^2 + \frac{1}{(n+1)^2}  \left \Vert b_n\right\Vert_2^2\nonumber\\
&& -  \frac{2}{n+1} Q_n^\top \mathbf{B}^{\frac{1}{2}}\pmb\theta_n - \frac{2}{n+1} b_n^\top \mathbf{B}^{\frac{1}{2}}\pmb\theta_n + \frac{2}{(n+1)^2} Q_n^\top b_n. \label{chenbeq}
\end{eqnarray}

From (\ref{cheneq2}), we have 
\begin{eqnarray} \label{chenvar}
\lim_{n \to \infty} (n+1)\Sigma_{n+1} = \mathbf{B}^{-1}. 
\end{eqnarray}
Define the Borel sigma-algebra
\begin{eqnarray*}
\mathcal{F}_n \triangleq \mathcal{B}\left(\bm x_{1}, ..., \bm x_{n+1}, \pmb\theta_1, ...., \pmb\theta_n, \tau_1, ..., \tau_{n+1}, y_1, ..., y_n, \delta_1, ..., \delta_n, \Sigma_1,..., \Sigma_n\right).
\end{eqnarray*}
Since $y_{n+1}$ is normally distributed and $\sup_x \left|\frac{d}{dx} \frac{\phi(x)}{1 - \Phi(x)}\right| \leq 1$, by (\ref{chenvar}) and Assumptions \ref{chena1}-\ref{chena2}, there must exist a positive constant $C_1$ such that for all $n$,  
\begin{eqnarray} \label{chens1}
\mathrm{E} \left(\left \Vert Q_n \right\Vert_2^2 | \mathcal{F}_n\right) &\leq& C_1 \left(1 +  \left\Vert \mathbf{B}^{\frac{1}{2}}\pmb\theta_n\right\Vert_2^2\right).
\end{eqnarray}
Similarly, with triangular inequality, there must also be a constant $C_2$ such that 
\begin{eqnarray} \label{chens2}
\mathrm{E} \left(\left \Vert b_n \right\Vert_2^2 | \mathcal{F}_n\right) &\leq& C_2 \left(1 +  \left\Vert \mathbf{B}^{\frac{1}{2}}\pmb\theta_n\right\Vert_2^2\right).
\end{eqnarray} 
By Cauchy-Schwarz inequality, from (\ref{chens1}) and (\ref{chens2}), we have 
\begin{eqnarray} 
\mathrm{E} \left(\left|2Q_n^\top b_n\right|  | \mathcal{F}_n\right) &\leq& \mathrm{E} \left(2 \left \Vert Q_n \right\Vert_2 \left \Vert b_n \right\Vert_2 | \mathcal{F}_n\right)\nonumber\\
&\leq& \mathrm{E} \left(\left \Vert Q_n \right\Vert_2^2 + \left \Vert b_n \right\Vert_2^2 | \mathcal{F}_n\right)\nonumber\\
&\leq& (C_1 + C_2)  \left(1 +  \left\Vert \mathbf{B}^{\frac{1}{2}}\pmb\theta_n\right\Vert_2^2\right). \label{chens3}
\end{eqnarray}
We can also find that 
\begin{eqnarray*}
\mathrm{E}\left(\left|\frac{2}{n+1} b_n^\top \mathbf{B}^{\frac{1}{2}}\pmb\theta_n\right| | \mathcal{F}_n\right) &\leq& \mathrm{E}\left(2 \left\Vert\frac{1}{(n+1)^{3/8}} b_n \right\Vert_2 \left\Vert \frac{1}{(n+1)^{5/8}} \mathbf{B}^{\frac{1}{2}}\pmb\theta_n\right\Vert_2 | \mathcal{F}_n\right)\\
&\leq& \mathrm{E}\left( \left\Vert\frac{1}{(n+1)^{3/8}} b_n \right\Vert_2^2 +  \left\Vert \frac{1}{(n+1)^{5/8}} \mathbf{B}^{\frac{1}{2}}\pmb\theta_n\right\Vert_2^2 | \mathcal{F}_n\right)\\
&\leq& \frac{1}{(n+1)^{3/4}} \mathrm{E}\left(\left\Vert b_n \right\Vert_2^2 | \mathcal{F}_n\right) +  \frac{1}{(n+1)^{5/4}} \left\Vert  \mathbf{B}^{\frac{1}{2}}\pmb\theta_n\right\Vert_2^2, 
\end{eqnarray*}
where the first inequality holds by Cauchy-Schwarz inequality. Since $y_{n+1}$ is normally distributed and $\sup_x \left|\frac{d}{dx} \frac{\phi(x)}{1 - \Phi(x)}\right| \leq 1$, by (\ref{chenvar}) and Assumptions \ref{chena1}-\ref{chena2}, there must exist two positive constants $C_3$ and $C_4$ such that 
\begin{eqnarray*}
\mathrm{E}\left(\left\Vert b_n \right\Vert_2^2 | \mathcal{F}_n\right) &\leq& C_3 \left\Vert (n+1)\Sigma_{n+1} - \mathbf{B}^{-1} \right\Vert_2^2 \left(1 + \left\Vert  \mathbf{B}^{\frac{1}{2}}\pmb\theta_n\right\Vert_2^2\right)\\
&=& C_3  \left \Vert (n+1)\Sigma_{n+1} \left(\frac{1}{n+1}\Sigma_{n+1}^{-1} -\mathbf{B}\right) \mathbf{B}^{-1}\right\Vert_2^2 \left(1 + \left\Vert  \mathbf{B}^{\frac{1}{2}}\pmb\theta_n\right\Vert_2^2\right)\\
&\leq& C_4 \left \Vert\frac{1}{n+1}\Sigma_{n+1}^{-1} -\mathbf{B} \right\Vert_2^2 \left(1 + \left\Vert  \mathbf{B}^{\frac{1}{2}}\pmb\theta_n\right\Vert_2^2\right),
\end{eqnarray*}

where the last inequality holds due to (\ref{chenvar}) and the submultiplicativity of the norm $ \left \Vert\cdot\right\Vert_2$. Thus, we have 
\begin{eqnarray}
\mathrm{E}\left(\left|\frac{2}{n+1} b_n^\top \mathbf{B}^{\frac{1}{2}}\pmb\theta_n\right| | \mathcal{F}_n\right) &\leq& \left( \frac{C_4 }{(n+1)^{3/4}} \left \Vert\frac{1}{n+1}\Sigma_{n+1}^{-1} -\mathbf{B} \right\Vert_2^2  + \frac{1}{(n+1)^{5/4}}\right) \left\Vert  \mathbf{B}^{\frac{1}{2}}\pmb\theta_n\right\Vert_2^2\nonumber\\
&& + \frac{C_4 }{(n+1)^{3/4}} \left \Vert\frac{1}{n+1}\Sigma_{n+1}^{-1} -\mathbf{B} \right\Vert_2^2. \label{chens4}
\end{eqnarray}

Finally, for any $\pmb\zeta$, we have 
\begin{eqnarray*}
\lefteqn{\mathrm{E} \left(Q_n^\top \mathbf{B}^{\frac{1}{2}}\pmb\zeta | \mathcal{F}_n\right)}\\
&=&  \bm x^\top_{n+1}\pmb\zeta \left(\frac{\bm x^\top_{n+1}\pmb\zeta}{\sigma^2} \Phi\left(\frac{\log \tau_{n+1}}{\sigma}\right) + \frac{1}{\sigma} \phi\left(\frac{\log \tau_{n+1}}{\sigma}\right) - \frac{1 - \Phi\left(\frac{\log \tau_{n+1}}{\sigma}\right)}{\sigma} \frac{\phi\left(\frac{\log \tau_{n+1} -  \bm x^\top_{n+1}\pmb\zeta}{\sigma}\right)}{1 - \Phi\left(\frac{\log \tau_{n+1} -  \bm x^\top_{n+1}\pmb\zeta}{\sigma}\right)}\right).
\end{eqnarray*}

Denote 
\begin{eqnarray*}
R_n\left(\bm x^\top_{n+1}\pmb\zeta\right) &\triangleq& \frac{\bm x^\top_{n+1} \pmb\zeta}{\sigma^2} \Phi\left(\frac{\log \tau_{n+1}}{\sigma}\right) + \frac{1}{\sigma} \phi\left(\frac{\log \tau_{n+1}}{\sigma}\right) - \frac{1 - \Phi\left(\frac{\log \tau_{n+1}}{\sigma}\right)}{\sigma} \frac{\phi\left(\frac{\log \tau_{n+1} -  \bm x^\top_{n+1}\pmb\zeta}{\sigma}\right)}{1 - \Phi\left(\frac{\log \tau_{n+1} -  \bm x^\top_{n+1}\pmb\zeta}{\sigma}\right)}, 
\end{eqnarray*}
then we have 
\begin{eqnarray}
\mathrm{E} \left(Q_n^\top \mathbf{B}^{\frac{1}{2}}\pmb\zeta | \mathcal{F}_n\right) =  \bm x^\top_{n+1}\pmb\zeta R_n\left(\bm x^\top_{n+1}\pmb\zeta\right). \label{chens5}
\end{eqnarray} 
Since $\left(\log \tau_n\right)_{n=0}^\infty$ is bounded and $\frac{d}{d u} R_n (u) >0$, we can see that $R_n\left(u\right) = 0$ if and only if $u = 0$, and for all $\epsilon > 0$,
\begin{eqnarray*}
\inf_{\left(\bm x_n^\top \pmb\zeta \right)^2 > \epsilon, n \in \mathbb{N}} \bm x_{n+1}^\top \pmb\zeta R_n\left(\bm x_{n+1}^\top \pmb\zeta\right) > 0.
\end{eqnarray*} 

Now, combining (\ref{chenbeq}) with (\ref{chens1}) - (\ref{chens5}), we have 
\begin{eqnarray*}
\lefteqn{\mathrm{E}\left(\left\Vert \mathbf{B}^{\frac{1}{2}}\pmb\theta_{n+1}\right\Vert_2^2 | \mathcal{F}_n\right)}\\
 &\leq&  \left\Vert \mathbf{B}^{\frac{1}{2}}\pmb\theta_n\right\Vert_2^2 \left(1 + \frac{2(C_1 + C_2)}{(n+1)^2} +  \frac{C_4 }{(n+1)^{3/4}} \left \Vert\frac{1}{n+1}\Sigma_{n+1}^{-1} -\mathbf{B} \right\Vert_2^2  + \frac{1}{(n+1)^{5/4}}\right)\\
&& + \frac{2(C_1 + C_2)}{(n+1)^2} + \frac{C_4 }{(n+1)^{3/4}} \left \Vert\frac{1}{n+1}\Sigma_{n+1}^{-1} -\mathbf{B} \right\Vert_2^2 -  \frac{2}{n+1}  \bm x^\top_{n+1}\pmb\theta_n R_n\left(\bm x^\top_{n+1}\pmb\theta_n\right).
\end{eqnarray*}
From Lemma \ref{chenl1}, we have 
\begin{eqnarray*}
\sum_{n=0}^\infty \frac{2(C_1 + C_2)}{(n+1)^2} +  \frac{C_4 }{(n+1)^{3/4}} \left \Vert\frac{1}{n+1}\Sigma_{n+1}^{-1} -\mathbf{B} \right\Vert_2^2  + \frac{1}{(n+1)^{5/4}} &<& \infty, \\
\sum_{n=0}^\infty \frac{2(C_1 + C_2)}{(n+1)^2} + \frac{C_4 }{(n+1)^{3/4}} \left \Vert\frac{1}{n+1}\Sigma_{n+1}^{-1} -\mathbf{B} \right\Vert_2^2 &<& \infty. 
\end{eqnarray*}
Then, by Theorem 1 in \cite{Robbins1985} , $\lim_{n \to \infty} \left\Vert \mathbf{B}^{\frac{1}{2}}\pmb\theta_n\right\Vert_2^2 $ exists and 
\begin{eqnarray*}
\sum_{n=0}^\infty \frac{1}{n+1} \bm x_{n+1}^\top \pmb\theta_n R_n(\bm x_{n+1}^\top\pmb\theta_n)  < \infty
\end{eqnarray*}
almost surely. Therefore, for every sample path, there must exist a subsequence $\left(\bm x_{n_k+1}^\top \pmb\theta_{n_k} \right)$ of $\left(\bm x_{n+1}^\top \pmb\theta_n\right)$ such that as $k \to \infty$, 
\begin{eqnarray*}
\bm x_{n_k+1}^\top \pmb\theta_{n_k} \to 0.
\end{eqnarray*}
On the other hand, since $\lim_{n \to \infty} \left\Vert \mathbf{B}^{\frac{1}{2}}\pmb\theta_n\right\Vert_2^2 $ exists, then for one sample path, the sequence $\left(\pmb\theta_n\right)$ is bounded. Therefore, there must exist a subsequence $\left(\pmb\theta_{n_{k_j}} \right)$ of $\left(\pmb\theta_{n_k} \right)$ such that as $j \to \infty$,
\begin{eqnarray*}
\pmb\theta_{n_{k_j}} \to \pmb\nu,
\end{eqnarray*}
where $\pmb\nu$ is a fixed vector. Then by Assumption \ref{chena2}, we have 
\begin{eqnarray*}
\lim_{j \to \infty} \left|\bm x_{n_{k_j}+1}^\top  \pmb\nu \right|  &=&  \lim_{j \to \infty} \left|\bm x_{n_{k_j}+1}^\top \left(\pmb\nu -\pmb\theta_{n_{k_j}} + \pmb\theta_{n_{k_j}}\right)\right|\\
&\leq&  \lim_{j \to \infty} \left|\bm x_{n_{k_j}+1}^\top \left(\pmb\nu -\pmb\theta_{n_{k_j}} \right)\right| +  \lim_{j \to \infty} \left|\bm x_{n_{k_j}+1}^\top \pmb\theta_{n_{k_j}}\right|\\
&=& 0.
\end{eqnarray*}
Thus, for any arbitrary $\epsilon > 0$, there exists an integer $J$ such that for all $j  \geq J$,
\begin{eqnarray} \label{chennu}
 \left|\bm x_{n_{k_j}+1}^\top \pmb\nu \right| < \epsilon.
\end{eqnarray}
However, since $\left(\bm x_{n_{k_j}+1} \right)_{j= J}^\infty$ is also an infinite sequence of i.i.d. samples from a common distribution, there must exist $K$ linearly independent vectors $\bm x_{n_{k_{j_1}}+1}, ..., \bm x_{n_{k_{j_K}}+1}$ from $\left(\bm x_{n_{k_j}+1} \right)_{j= J}^\infty$ that can be a basis of $\mathbb{R}^K$; otherwise, suppose all $\left(\bm x_{n_{k_j}+1} \right)_{j= J}^\infty$ come from a subspace $V$ of $\mathbb{R}^K$ and $V \neq \mathbb{R}^K$, then there must be a nonzero vector $\pmb\gamma \in V^{\bot}$ such that 
\begin{eqnarray*}
\pmb\gamma^\top \mathbf{A} \pmb\gamma &=& \pmb\gamma^\top \left(\lim_{J' \to \infty} \frac{1}{J'} \sum_{j=J}^{J'} \bm x_{n_{k_j}+1} \bm x_{n_{k_j}+1}^\top\right) \pmb\gamma\\
&=& \lim_{J' \to \infty} \frac{1}{J'} \sum_{j=J}^{J'} \left(\bm x_{n_{k_j}+1}^\top \pmb\gamma\right)^2\\
&=& 0,
\end{eqnarray*}
where the first equality holds by Assumptions \ref{chena1}-\ref{chena2}, but this contradicts Assumption \ref{chena1} that $\mathbf{A}$ is positive-definite.

Then, to satisfy (\ref{chennu}), since $\epsilon$ can be arbitrarily small, by Assumption \ref{chena2}, $\pmb\nu$ has to be the zero vector. Thus, $\pmb\theta_{n_{k_j}} \to 0$, so $
\lim_{j \to \infty} \left\Vert \mathbf{B}^{\frac{1}{2}}\pmb\theta_{n_{k_j}}\right\Vert_2^2 = 0$,
but $\left(\pmb\theta_{n_{k_j}}\right)$ is a subsequence of $\left(\pmb\theta_n\right)$ and $\lim_{n \to \infty} \left\Vert \mathbf{B}^{\frac{1}{2}}\pmb\theta_n\right\Vert_2^2 $ exists; therefore, $\lim_{n \to \infty} \left\Vert \mathbf{B}^{\frac{1}{2}}\pmb\theta_n\right\Vert_2^2 = 0$, so we have $\pmb\theta_n \to 0$ for every sample path, thus $\pmb\theta_n \to 0$ almost surely.

\end{document}